\documentstyle[prl,aps]{revtex}        
\input epsf
\begin{document}
\draft
\twocolumn
\title{Dielectric formalism and damping of collective modes
in trapped Bose-Einstein condensed gases} 
\author{Gyula Bene$^{(1)}$ and P\'eter Sz\'epfalusy$^{(2,3)}$}
\address{$^{(1)}$ Institute for Theoretical Physics, E\"otv\"os University,
Puskin u. 5-7, H-1088 Budapest, Hungary, $^{(2)}$ Department of Physics of Complex Systems,
E\"otv\"os University,
M\'uzeum k\"or\'ut 6-8, H-1088 Budapest, Hungary and
$^{(3)}$ Research Institute for Solid State Physics and Optics
of the Hungarian Academy of Sciences, P.O.Box 49, H-1525 Budapest, Hungary}
\date{\today}
\maketitle


\begin{abstract}
We present the general dielectric formalism for Bose-Einstein condensed 
systems in external potential at finite temperatures. On the basis 
of a model arising within this framework as a first approximation
in an intermediate temperature region for large condensate
we calculate the damping of low-energy excitations in the collisionless
regime.
\end{abstract}
\narrowtext
The recent experimental observation of Bose-Einstein condensation in alkali metal vapours 
\cite{Anderson}, \cite{Bradley} , \cite{Davis}
gave a new impetus to the research of degenerate Bose systems. Much of the theoretical
apparatus developed in the sixties for the study of superfluid helium 
can be well applied in this case also. The present situation is even a bit simpler,
as, due to the low density, interatomic collisions play a much less important role
than in case of the helium II.
There is, however, an additional complication due to the presence of the external
trapping potential.

The Bose-Einstein condensation is also a prime example 
of continuous phase transitions and spontaneous symmetry
breaking.
Indeed, the presence of the condensate implies that
 the phase invariance of the Hamiltonian is absent in the 
ground state of the system.
One of the most spectacular consequences of a broken continuous symmetry
is the strong and direct coupling among different correlation (response) functions.
Its consequences have been most fully exploited by the dielectric
formalism first introduced at zero temperature \cite{Ma}, \cite{Kondor-Szepfalusy}
and later generalized to finite temperature\cite{Szepfalusy-Kondor}, \cite{Payne-Griffin}
(for an enlightening discussion of features and further references see \cite{Griffin}). 
A most useful scheme for calculation has been 
developed by Gould and Wong\cite{Gould-Wong}. Note that the formalism
proved to be useful in interpreting neutron scttering
experiments in liquid helium (see papers by H.R.Glyde and by A.Griffin in \cite{exc} and for
a recent experiment \cite{Blago}).

Our purpose is to generalize the dielectric formalism to inhomogeneous systems
and apply it to trapped Bose condensed gases. The emphasis
in this paper besides general considerations is put on a simple model emerging 
from the general theory and the calculation of the damping
of the lowest lying excitations in the collisionless regime. This is motivated by the fact that
at present the measured damping rates of low frequency collective excitations \cite{JILA1}-\cite{MIT2}
are in the focus
of interest, partly, because they are more characteristic of the Bose-Einstein
condensation than the excitation frequencies.
Recently a number of papers have been devoted to the calculation of the damping 
in the collisionless regime
in case of homogeneous systems and of trapped gases as well \cite{Liu}-\cite{Fedichev-Shlyapnikov2}.
In particular the results in \cite{Fedichev-Shlyapnikov} are directly relevant to the experimental
situation \cite{JILA2}.

Let us start by summarizing the extension of the dielectric formalism to the inhomogeneous system.
This formalism emphasizes the importance of quantities 
which are proper and irreducible simultaneously. 
Having in mind perturbation theory 
and the corresponding Feynman graphs,
we shall call the proper part of a quantity a part
which cannot be cut into two by splitting
a single interaction line. In contrast, an irreducible part
will be a graph which cannot be cut into two by splitting
a single line corresponding to a bare propagator 
(i.e., unperturbed Green's function).

As usually, we start from the Hamiltonian 
\begin{eqnarray}
\hat H=\int d^3 \vec{r} \hat \psi^\dagger(\vec{r})\left(-\frac{\nabla^2}{2m}+U(\vec{r})\right)
\hat \psi(\vec{r})\nonumber\\
+\frac{1}{2}\int d^3 \vec{r_1}\int d^3 \vec{r_2}
 \hat \psi^\dagger(\vec{r_1}) \hat \psi^\dagger(\vec{r_2})v(\vec{r_1},\vec{r_2})
 \hat \psi(\vec{r_1}) \hat \psi(\vec{r_2})\label{ham}
\end{eqnarray}
assuming two-body interactions. Here $\hat \psi(\vec{r})$ is the Bose field operator
and $U(\vec{r})$ denotes the external potential.
Separating out the condensate part we have 
$\hat \psi(\vec{r})=\Phi_0(\vec{r})+\hat \phi(\vec{r})$, where
$\Phi_0(\vec{r})=<\hat \psi(\vec{r})>$ stands for the condensate wave function.
The angular bracket here and hereafter denotes ensemble averaging with the grand canonical
statistical operator $e^{-\beta(\hat H -\mu\hat N)}/{\rm Tr} e^{-\beta(\hat H -\mu\hat N)}$.
The density autocorrelation function is defined by 
$\chi(\vec{r_1},\tau,\vec{r_2},\tau')=
-\left< T_\tau \left[\tilde n(\vec{r},\tau)\tilde n(\vec{r'},\tau')\right]\right>$,
where $\tilde n(\vec{r})=\hat n(\vec{r})-\left<\hat n(\vec{r})\right>
=\Phi_0(\vec{r})\hat \phi^\dagger(\vec{r}) + \Phi_0(\vec{r})\hat \phi(\vec{r})
+\hat \phi^\dagger(\vec{r})\hat \phi(\vec{r})-\left<\hat \phi^\dagger(\vec{r})\hat \phi(\vec{r})\right>$
is the density deviation operator.
The field operators here are given in Matsubara representation, $\tau$
playing the role of imaginary time.
We shall also need the temperature Green's functions defined by $G_{1,1}(\vec{r},\tau,\vec{r'},\tau')=
G_{2,2}(\vec{r'},\tau',\vec{r},\tau)=
-\left< T_\tau \left[\hat \phi(\vec{r},\tau)\hat \phi^\dagger(\vec{r'},\tau')\right]\right>$,\hfill\break 
$G_{1,2}(\vec{r},\tau,\vec{r'},\tau')=
-\left< T_\tau \left[\hat \phi(\vec{r},\tau)\hat \phi(\vec{r'},\tau')\right]\right>$\hfill\break  and 
$G_{2,1}(\vec{r},\tau,\vec{r'},\tau')$ is obtained by replacing $\hat \phi$ by $\hat \phi^\dagger$ 
in $G_{1,2}(\vec{r},\tau,\vec{r'},\tau')$.
In the Fourier transforms 
$G_{\alpha,\beta}(\vec{r},\vec{r'},\omega)$
and $\chi(\vec{r_1},\vec{r_2},\omega)$
the frequency
$\omega$  takes on only the discrete values $\omega_n=2n\pi/(\hbar \beta)$
where $n$ is an integer. 
The physically meaningful retarded correlation function and Green's function
can be constructed by
a suitable analytic continuation\cite{Fetter-Walecka}. We shall denote 
both the temperature and the retarded functions by the same symbol.

The proper part of a quantity (correlation function, Green's function etc.)
will be denoted by a tilde over the corresponding symbol.
The definition of the proper part directly implies that
\begin{eqnarray}
\chi(\vec{r_1},\vec{r_2},\omega)
=\tilde\chi(\vec{r_1},\vec{r_2},\omega)\mbox{\hspace{4cm}}\nonumber\\
+\frac{1}{\hbar}\int d^3 \vec{r_3} 
\int d^3 \vec{r_4} \,\tilde \chi(\vec{r_1},\vec{r_4},\omega)
v(\vec{r_4},\vec{r_3})\, \chi(\vec{r_3},\vec{r_2},\omega)\,,\label{e1}
\end{eqnarray}
Eq.(\ref{e1}) remains valid also for the corresponding retarded correlation
functions (i.e., after analytic continuation). When $\chi(\vec{r_3},\vec{r_2},\omega)$
has a pole in $\omega$ (and $\tilde\chi(\vec{r_3},\vec{r_2},\omega)$ is nonsingular
at that $\omega$ value), 
\begin{eqnarray}
\int d^3 \vec{r_3} 
\left(\delta(\vec{r_1}-\vec{r_3})-\frac{1}{\hbar}
\int d^3 \vec{r_4} \tilde \chi(\vec{r_1},\vec{r_4},\omega)
v(\vec{r_4},\vec{r_3})\right)\nonumber\\\times \xi(\vec{r_3})=0\label{e2}
\end{eqnarray}
should hold, where $\xi(\vec{r_3})$ is an eigenfunction.

A remarkable feature of the Bose condensed system is the 'mixing'
of correlation functions of different order, that is 
a consequence of the related symmetry breaking 
\cite{Ma}-\cite{Griffin}.
This is the basis of the dielectric formalism and
can be expressed by the generalization of one of its basic relationships  
to inhomogeneous systems as
\begin{eqnarray}
\tilde\chi(\vec{r_1},\vec{r_2},\omega)
=\tilde\chi^{(r)}(\vec{r_1},\vec{r_2},\omega)\mbox{\hspace{3cm}}\nonumber\\
+\int d^3 \vec{r_3} \int d^3 \vec{r_4} \tilde \Lambda_\alpha (\vec{r_1},\vec{r_3},\omega)\,
\tilde G_{\alpha,\beta} (\vec{r_3},\vec{r_4},\omega)\label{e3}\\\times
 \Lambda_\beta (\vec{r_4},\vec{r_2},\omega)\,.\nonumber
\end{eqnarray}
Here $\tilde\chi^{(r)}$ is the proper and irreducible regular part of the density 
correlation function,  $\tilde \Lambda_\alpha$ stands for the proper part of the anomalous vertex
$\Lambda_\alpha$ which
is due to the presence of the condensate and represents the contribution
of graphs with one outer particle line and one outer interaction line. 
$\tilde G_{\alpha,\beta}$ is the proper part of the
Green's function. For repeated 'spinor' indices a summation is understood.

Eq.(\ref{e3}) allows us to derive the relation
\begin{eqnarray}
\int d^3 \vec{r_2}\int d^3 \vec{r_3}\tilde \Lambda_\gamma(\vec{r_1},\vec{r_2},\omega)
\tilde G_{\gamma,\beta}(\vec{r_2},\vec{r_3},\omega)
G^{-1}_{\beta,\alpha}(\vec{r_3},\vec{r_4},\omega)\nonumber\\
=
\int d^3 \vec{r_2}\int d^3 \vec{r_3}
\tilde \chi(\vec{r_1},\vec{r_2},\omega)
\chi^{-1}(\vec{r_2},\vec{r_3},\omega)\Lambda_\alpha(\vec{r_3},\vec{r_4},\omega)\label{f5}
\end{eqnarray}
Here the inverses are understood in the integral operator sense.
This equation implies the coincidence of the eigenvalues of the
density correlation functions and the Green's functions. The 
eigenvalue equation in case of the Green's functions is
\begin{eqnarray}
\frac{1}{\hbar}
\int d^3 \vec{r_4} \tilde G_{\alpha,\beta}(\vec{r_1},\vec{r_4},\omega)
\mbox{\hspace{4.5cm}}\nonumber\\\times 
\left(\Sigma_{\beta,\gamma}(\vec{r_4},\vec{r_3},\omega)
-\tilde \Sigma_{\beta,\gamma}(\vec{r_4},\vec{r_3},\omega)\right)
\varphi_\gamma(\vec{r_3})=\varphi_\alpha(\vec{r_1})\,.\label{ee2}
\end{eqnarray}
Here $\Sigma_{\beta,\gamma}$ stands for the (irreducible)
self-energy, and  $\tilde \Sigma_{\beta,\gamma}$ for its proper part.
The eigenvectors $\xi$ and $\varphi_\alpha$ belonging to the same eigenvalue $\omega$ 
are related by
$\xi(\vec{r_1})=\int d^3 \vec{r_2}\Lambda_\alpha(\vec{r_1},\vec{r_2},\omega)
 \varphi_\alpha(\vec{r_2})$,
or, the other way round,
\begin{eqnarray}
\varphi_\alpha(\vec{r_1})=\int d^3 \vec{r_2}\int d^3 \vec{r_3}\int d^3 \vec{r_4}
\tilde G_{\alpha,\beta}(\vec{r_1},\vec{r_2},\omega)\nonumber\\
\times \tilde \Lambda_\beta(\vec{r_2},\vec{r_3},\omega)
v(\vec{r_3},\vec{r_4})
 \xi(\vec{r_4})\,.\label{f8}
\end{eqnarray}

Up to now we have given expressions that are valid in all order of the perturbation
theory. Note that in the Bogolyubov approximation Eq.(\ref{ee2}) can be transformed into the
familiar Bogolyubov equations for $u$ and $-v$, the components of
$\varphi_\alpha$. The same remains true in the finite
temperature generalization of the Bogolyubov theory by Popov \cite{Popov}, \cite{Griffin}.

We turn now to present a model based upon the general formalism introduced above and to 
work out the damping of the low energy excitations in this framework in the weak-coupling
regime $<\hat n>g << k_B T$. Here $g=4\,\pi\,\hbar^2\,a/m$, where $a$ is the $s$-wave
scattering length assumed to be positive and to describe the interaction between atoms at the
relevant energies.
The model arising
is the generalization to nonzero external potential that of treated in detail
for homogeneous system in \cite{Szepfalusy-Kondor}.

The anomalous vertex $\Lambda_\alpha$ will be approximated by
\begin{eqnarray}
\Lambda_\alpha(\vec{r_1},\vec{r_2},\omega)=\Phi_0(\vec{r_1})
\delta(\vec{r_1}-\vec{r_2})\,,\label{e6}
\end{eqnarray}
where the condensate wave function $\Phi_0(\vec{r_1})$ is assumed to be real.

As for the proper Green's functions, 
we shall consider the approximation when 
\begin{eqnarray}
\tilde G_{1,2}=\tilde G_{2,1}=0\label{e4a}\\
\left(\hbar\omega-\tilde H\right)
\tilde G_{1,1}(\vec{r_1},\vec{r_2},\omega)
=\hbar\delta(\vec{r_1}-\vec{r_2})\,.\label{e4}
\end{eqnarray}
Here 
$\tilde H=\hat T + U(\vec{r_1})
-\mu+g\Phi_0^2(\vec{r_1})+2\,g\,n_T(\vec{r_1})$
 with  $\hat T=-\frac{\hbar^2}{2\,m}\Delta$ standing 
for the operator of the kinetic energy and $\mu$ for the chemical potential.
Further, $n_T(\vec{r})=\left<\hat \phi^\dagger(\vec{r})\hat \phi(\vec{r})\right>$ is the
density of the noncondensate part. 
Note that 
$\tilde G_{2,2}(\vec{r_1},\vec{r_2},\omega)=\tilde G_{1,1}(\vec{r_2},\vec{r_1},-\omega)$.
Eq.(\ref{e4}) can be derived when the Hartree-Fock terms are retained in the 
self-energy graphs for the contribution of thermally excited particles, which 
corresponds to the Popov approximation \cite{Popov}, \cite{Griffin}.

Inserting Eqs.(\ref{e3}),
(\ref{e6}) into Eq.(\ref{e2})
we get
\begin{eqnarray}
\xi(\vec{r_1})-\frac{g}{\hbar}\int d^3 \vec{r_2} 
\tilde \chi^{(r)}(\vec{r_1},\vec{r_2},\omega)\xi(\vec{r_2})\nonumber\\
-\frac{g}{\hbar}\phi_0(\vec{r_1})\int d^3 \vec{r_2} 
\left[\tilde G_{1,1}(\vec{r_1},\vec{r_2},\omega)\right.\nonumber\\\left.
+\tilde G_{1,1}(\vec{r_2},\vec{r_1},-\omega)\right]
\phi_0(\vec{r_2})\xi(\vec{r_2})=0\,.\label{e8}
\end{eqnarray}
Let us divide this equation by $\phi_0(\vec{r_1})$ and apply to it the operator
$
\left(\hbar\omega-\tilde H\right)\left(-\hbar\omega-\tilde H\right)$.

We get
\begin{eqnarray}
2\,g\tilde H\Phi_0(\vec{r_1})\xi(\vec{r_1
})=\left(\hbar^2\omega^2-\tilde H^2\right)\mbox{\hspace{3cm}}\nonumber\\
\times
\left(\frac{\xi(\vec{r_1})}{\Phi_0(\vec{r_1})}-\frac{g}{\hbar}\frac{1}{\Phi_0(\vec{r_1})}
\int d^3 \vec{r_2} \tilde \chi^{(r)}(\vec{r_1},\vec{r_2},\omega)
\xi(\vec{r_2})\right)\label{e9}
\end{eqnarray}

The condensate wave function $\Phi_0(\vec{r})$ can be determined
from the Gross-Pitaevskii equation \cite{Fetter-Walecka}
\begin{eqnarray}
\tilde H\Phi_0(\vec{r})=0
\label{egp}
\end{eqnarray}

Keeping only the term containing $\omega^2$ on the r.h.s. of Eq.(\ref{e9}), 
neglecting $\tilde \chi^{(r)}$ and using Eq.(\ref{egp}), we arrive at 

\begin{eqnarray}\omega_0^2\,\xi_0(\vec{r})
=-\frac{g}{m}\vec{\nabla} \left(
\left(\Phi_0(\vec{r})\right)^2\,\vec{\nabla}\xi_0(\vec{r})\right)
\,.\label{e12}
\end{eqnarray}
This amounts to the hydrodynamic approximation  known to yield accurate
values for the excitation frequences when $\hbar\omega << \mu$ \cite{hydro1}, 
\cite{hydro2}, \cite{Ohberg}. Hereafter we shall be interested in this regime, and 
our aim is the calculation of the decay rate. This is determined in leading order by
the term containing $\tilde \chi^{(r)}$ in Eq.(\ref{e9}). Therefore, we consider this term
a perturbation and solve Eq.(\ref{e9}) perturbatively (still omitting
the small term containing $\tilde H^2$ which does not influence the decay
to this order). 

We write the frequency as
$\omega=\omega_0+\omega_1$
where the correction $\omega_1$ is supposed to be small compared to the
leading term $\omega_0$. Similarly, the eigenfunction is decomposed as
$\xi(\vec{r})=\xi_0(\vec{r})+\xi_1(\vec{r})$.

Collecting the corrections we get in first order
\begin{eqnarray}
2\,\omega_0\,\omega_1 \xi_0(\vec{r_1})
-\omega_0^2
\frac{g}{\hbar}
\int d^3 \vec{r_2} \tilde \chi^{(r)}(\vec{r_1},\vec{r_2},\omega_0)
\xi_0(\vec{r_2})\nonumber\\
=-\omega_0^2\,\xi_1(\vec{r})-\frac{g}{m}\vec{\nabla} \left(
\left(\Phi_0(\vec{r_1})\right)^2\,\vec{\nabla}\xi_1(\vec{r_1})\right)
\,.\label{e13}
\end{eqnarray}
Multiplying Eq.(\ref{e13}) with $\xi_0(\vec{r_1})$ and integrating over 
$\vec{r_1}$ the r.h.s. of the equation identically vanishes
(as the operator appearing in the hydrodynamic equation is a self-adjoint one),
and we are left with an equation not containing
$\xi_1(\vec{r})$ any longer, i.e., with an equation for $\omega_1$ alone.
It gives
\begin{eqnarray}
\omega_1
=\frac{g\,\omega_0}{2\,\hbar}
\frac{\int d^3 \vec{r_1}\int d^3 \vec{r_2} \xi_0^*(\vec{r_1})\tilde\chi^{(r)}(\vec{r_1},\vec{r_2},\omega_0)
\xi_0(\vec{r_2})}
{\int d^3 \vec{r_1}|\xi_0(\vec{r_1})|^2}\,.\label{e17}
\end{eqnarray}

The damping is determined by the imaginary part of this expression.
When evaluating Eq.(\ref{e17}) we apply the Thomas-Fermi approximation
for the determination of $\Phi_0(\vec{r})$, as a relevant approximation
in the regime where experiments are done. This means that in Eq.(\ref{egp})
we may neglect the kinetic energy term compared to the others. 
Moreover, when solving (\ref{e12}), the finiteness of the temperature will be taken into account 
only in the normalization 
condition of the condensate wave function.
For a justification of
this procedure see Ref.\cite{Burnett}.
Thus we use $
\left(\Phi_0(\vec{r})\right)^2=\left(\mu_0-U(\vec{r})\right)/g$
with $N_0/N=1-(T/T_c)^3$ where $N$ and $N_0$ are the total and the condensate number of
particles, respectively.

By inserting $\Phi_0(\vec{r})$ into Eq.(\ref{e12})
the zeroth order eigenfunction $\xi_0(\vec{r})$ is determined \cite{hydro2}, \cite{Ohberg}
by separating Eq.(\ref{e12}) in oblate spherical coordinates and using a
polynomial Ansatz.

The quantity $\tilde\chi^{(r)}(\vec{r_1},\vec{r_2},\omega)$ is approximated by
the (dressed) loop $=\sum_n \tilde G_{\alpha,\beta}(\vec{r_1},\vec{r_2},\omega_n)
\tilde G_{\beta,\alpha}(\vec{r_2},\vec{r_1},\omega_n-\omega)$
where the proper Green's function $\tilde G_{\alpha,\beta}(\vec{r_1},\vec{r_2},\omega_n)$ satisfies 
Eqs.(\ref{e4a}) and (\ref{e4}). 
Note that according to (\ref{e17}) we need $\tilde\chi^{(r)}$ only inside
the condensate. In the Thomas-Fermi approximation 
applied to $\tilde G$ one obtains for large condensate
\begin{eqnarray}
\tilde\chi^{(r)}(\vec{r_1},\vec{r_2},\omega)=\frac{1}{(2\pi)^3}\int d^3\vec{k}
e^{i\vec{k}(\vec{r_1}-\vec{r_2})}\tilde \chi^{(r)}(\vec{k},\omega)\,.
\end{eqnarray}
where $\tilde\chi^{(r)}(\vec{k},\omega)$ stands for the bubble graph 
with free propagators investigated in \cite{Szepfalusy-Kondor}
in detail. This means that the spectrum
related to $\tilde G$ is quasicontinuous in our approximation.
For the damping we need only the imaginary part of 
$\tilde \chi^{(r)}(\vec{k},\omega)$, which can be calculated analytically \cite{Szepfalusy-Kondor}.

The expression (\ref{e17}) for the damping was evaluated (using the above listed
approximations) by Monte-Carlo integration, using $10^8$ points for
the numerator. The Fourier transform 
has previously been evaluated numerically. The error of the evaluation 
of the expression (\ref{e17}) was less than 1\%. 
The weak coupling condition is satisfied if $T/T_c > 0.6$ taking the parameters of the
JILA experiment \cite{JILA2}.
At temperatures close to $T_c$ the Thomas-Fermi approximation loses
its validity. Choosing as an intermediate temperature $T/T_c=0.7$ our result is
$118\,s^{-1}$ ($115\,s^{-1}$) for the damping rate at $m=0$ ($m=2$),
where $m$ is the value of the $L_z$ angular momentum component, 
according to which 
the solutions of Eq.(\ref{e12}) can be classified
due to the cylindrical symmetry. This agrees within experimental error with the measured value
\cite{JILA2} and also with the result of
\cite{Fedichev-Shlyapnikov} which was derived using approximations
different from ours. 

We have applied Eq.(\ref{e2}) with Eq.(\ref{e3}) in our calculation, i.e., 
treated the density fluctuations, which are directly measured. The Eqs.(\ref{f5})-(\ref{f8})
of the dielectric formalism make possible to determine the corresponding parameters 
(not presented here) of the
Green's functions $G_{\alpha,\beta}$ and the eigenfunctions $\varphi_\alpha$
when the temperature is below the critical one. A basic advantage of the dielectric formalism
is that the decoupling of the density and one-particle fluctuation spectra can be followed
when the temperature increases across $T_c$ \cite{Szepfalusy-Kondor}, a fact which has been
exploited in case of liquid helium \cite{exc}, \cite{Blago}. In case of trapped Bose gases further
experiments along these lines would help further developing the theory.

\newpage
{\bf Acknowledgements}
\vskip0.5cm

The present work has been partially supported by
the Hungarian National Research Foundation under Grant Nos. OTKA T017493 and
OTKA F 17166, 
 by the Hungarian Ministry of Culture and Education 
 under Grant No. FKFP 0159/1997.


\begin{thebibliography}{99}
\bibitem{Anderson} M.H.Anderson, J.R.Ensher, M.R.Matthews,\hfill\break C.E.Wiemann, and
E.A.Cornell,  Science {\bf 269}, 198 (1995).
\bibitem{Bradley} C.C.Bradley, C.A.Sacket, J.J.Tollett, and R.G.Hulet,
Phys.Rev.Lett. {\bf 75}, 1687 (1995).
\bibitem{Davis} K.B.Davis, M.O.Mewes, M.R.Andrews, N.J. van Drouten,
D.D.Durfee, D.M.Kurn, and W.Ketterle, Phys.Rev.Lett. {\bf 75}, 3969 (1995). 
\bibitem{Ma} Shang-Keng Ma and Chia-Wei Woo, Phys.Rev. {\bf 159}, 165 (1967).
\bibitem{Kondor-Szepfalusy} I.Kondor and P.Sz\'epfalusy, Acta Phys. Hung. {\bf 24}, 81 (1968).
\bibitem{Szepfalusy-Kondor} P.Sz\'epfalusy and I.Kondor, Ann. Phys. {\bf 82} (1974) 1.
\bibitem{Payne-Griffin} S.H.Payne and A.Griffin, 
Phys.Rev.B {\bf 32}, 7199 (1985).
\bibitem{Griffin} A.Griffin, {\em Excitations in a Bose-Condensed Liquid},
Cambridge University Press, 1993.
\bibitem{Gould-Wong}  V.K.Wong and H.Gould, Annals of Physics, {\bf 83},
252 (1974).
\bibitem{exc} {\em Excitations in Two-Dimensional and Three-Dimensional
Quantum Fluids}, ed. A.G.F.Wyatt and H.J.Lauter, Plenum Press, New York, 1991.
\bibitem{Blago} N.M.Blagoveshchenskii e.al., Phys.Rev. {\bf B 50}, 16550 (1994).
\bibitem{JILA1} D.S.Jin, J.R.Ensher, M.R.Matthews, C.E.Wieman, and E.A.Cornell,
Phys.Rev.Lett. {\bf 77}, 420 (1996).
\bibitem{MIT1} M.O.Mewes, M.R.Andrews, N.J. van Drouten, D.M.Kurn, 
C.G.Townsend, and W.Ketterle, Phys.Rev.Lett. {\bf 77}, 988 (1996).
\bibitem{JILA2} D.S.Jin, M.R.Matthews, J.R.Ensher, C.E.Wieman, and E.A.Cornell,
Phys.Rev.Lett. {\bf 78}, 764 (1997).
\bibitem{MIT2} D.M.Stamper-Kurn, H.-J.Miesner,S.Inouye,
M.R.Andrews and W.Ketterle, {\em Excitations of a Bose-Einstein Condensate at
Nonzero Temperature: A Study of Zeroth, First, and Second Sound}, 
cond-mat/9801262.
\bibitem{Liu} W.Vincent Liu, Phys.Rev.Lett. {\bf 79}, 4056 (1997).
\bibitem{Pitaevskii-Stringari} L.P.Pitaevskii and S.Stringari, Phys.Lett. {\bf A 235} (1997).
\bibitem{Fedichev-Shlyapnikov} P.O.Fedichev, G.V.Shlyapnikov and
J.T.M.Walraven, Phys.Rev.Lett. {\bf 80}, 2269 (1998).
\bibitem{HuaShi-Griffin} Hua Shi and A.Griffin,
{\em Finite Temperature Excitations in a Dilute bose-Condensed Gas},
Phys.Rep., to appear.
\bibitem{Giordini} S.Giordini, cond-mat/9709259.
\bibitem{Minguzzi} A.Minguzzi and M.P.Tosi, J.Phys.: Condens. Matter {\bf 9}, 10211 (1997).
\bibitem{Fedichev-Shlyapnikov2} P.O.Fedichev and G.V.Shlyapnikov, cond-mat/9805015. 
\bibitem{Fetter-Walecka} 
E.M.Lifshitz and L.P.Pitaevskii, Statistical Physics, Part 2, Pergamon
Press, Oxford, New York, 1980.
\bibitem{Popov} V.N.Popov, {\em Functional Integrals
and Collective Modes}, (Cambridge University Press, New York, 1987), Chap. 6.
\bibitem{hydro1} S.Stringari, Phys.Rev.Lett. {\bf 77}, 2360 (1996).
\bibitem{hydro2} M.Fliesser, A.Csord\'as, P.Sz\'epfalusy and R.Graham,
Phys.Rev. A {\bf 56}, R2533 (1997).
\bibitem{Ohberg} P.Ohberg, E.L.Surkov, I.Tittonen, S.Stenholm,
M.Wilkens and G.V.Shlyapnikov, Phys.Rev. {\bf A 56}, R3346 (1997).
\bibitem{Burnett} R.J.Dodd, M.Edwards, C.W.Clark, K.Burnett, 
Phys.Rev. {\bf A 57}, R32 (1998). 


\end{thebibliography}
\end{document}